\RequirePackage{fix-cm}
\documentclass[smallextended]{svjour3}       
\smartqed  
\usepackage{graphicx}
\usepackage[T1]{fontenc}

\usepackage{bm}
\usepackage{amsmath}
\usepackage{amsfonts}
\usepackage{amssymb}
\usepackage{hyperref}
\usepackage{listings}
\usepackage{cite}
\usepackage{multirow}

\usepackage{dsfont}
\usepackage{array}
\usepackage{wrapfig}
\usepackage[font=scriptsize,labelfont=bf]{caption}
\usepackage{subcaption}
\usepackage{rotating}
\usepackage{array}
\usepackage{todonotes}
\usepackage{lscape}
\usepackage{natbib}

\newcommand{\phe}{\bm{z}}
\newcommand{\fit}{\omega}
\newcommand{\trate}{\tau}

\newcolumntype{L}[1]{>{\raggedright\let\newline\\\arraybackslash\hspace{0pt}}m{#1}}
\newcolumntype{C}[1]{>{\centering\let\newline\\\arraybackslash\hspace{0pt}}m{#1}}
\newcolumntype{R}[1]{>{\raggedleft\let\newline\\\arraybackslash\hspace{0pt}}m{#1}}

\begin{document}

\title{Modelling the proliferation of transposable elements in populations under environmental stress.\thanks{The work was supported by National Science Centre,  grant no. 2012/06/M/ST6/00438}}

\author{Krzysztof Gogolewski         \and
        Micha\l{ } Startek \and
        Arnaud Le Rouzic$^{\star}$ \and 
        Anna Gambin$^{\star}$\thanks{$^{\star}$These authors share senior authorship} 
}

\institute{Krzysztof Gogolewski \at
              Institute of Informatics, University of Warsaw, Poland 
           \and
          Micha\l{ } Startek \at
              Institute of Informatics, University of Warsaw, Poland 
      	   \and
      	  Anna Gambin \at 
      	  	 Institute of Informatics, University of Warsaw, Poland 
      	   \and
      	  Arnaud Le Rouzic \at 
      	      \'Evolution, G\'enomes, Comportement, \'Ecologie, CNRS -- IRD -- Univ.~Paris-Sud, Universit\'e Paris-Saclay, Gif-sur-Yvette, France 
}

\date{Received: date / Accepted: date}

\maketitle

\begin{abstract}

Although transposable elements were discovered in the middle of the 20$^\text{th}$ 
century and are known as one of the main genome components, their impact on genome evolution has not been fully elucidated yet. 
However, it was so far shown that mutations induced by transposition events
can lead to significant genetic disorders and affect an organism's phenotype. Furthermore, 
since every mutation can also be beneficial, the natural question that arises is whether under certain
conditions the activity of TEs can be considered as an evolutionary helper i.e. a mechanism that has positive
contribution to the evolution of a population. \\
In this article, we investigate the evolution of sexual diploid populations which are 
hosts for active TE families. Our purpose is to explore the relationship between the environmental change, that 
influences such population and activity of those TEs that are present in genomes. 
Based on results obtained from the stochastic computational model, we conclude that in presence of
progressive environmental change the activity of TEs, and specifically mutations that are mediated 
by their activity, can noticeably facilitate the process of adaptation to varying conditions and 
prevent the population from extinction. 
\end{abstract}

\section{Introduction}
\label{ch:intro}

Current models in evolutionary quantitative genetics describe convincingly how populations and species respond to artificial or natural selection from their standing genetic variation \cite{LW97,Rof97,CC10}. However, deciphering the mechanisms involved in the evolutionary properties of the genetic variation itself is more challenging, up to the point that a theory of the evolution of evolvability is often presented as a major piece of a potential new evolutionary synthesis \cite{KG98,Bro01,Pig07,PM10}. 

Populations' evolvability is fuelled by mutations. The rate at which mutations occur is known to vary by several orders of magnitude across organisms \cite{DCC+98,Lyn07b,BMD07}. Furthermore, multiple independent examples of spontaneous evolution of different mutation rates in the wild or in lab conditions \cite{GMT+01,DM06} confirm that the mutation rate of an individual is a variable phenotypic trait susceptible to be targetted by natural selection. 

Most mutations being deleterious, individuals harboring a "mutator" allele leading to a high mutation rate are unlikely to be favored by natural selection. However, in a continuously variable environment, a nil mutation rate necessarily drives species to extinction. Population genetics theory has thus focused on estimating equilibrium mutation rates, resulting from a balance between natural selection against deleterious mutations, driving mutation rates downwards, and mechanisms pulling mutation rates upwards, including the metabolic cost of DNA proof-reading, and a hypothetical -- but outstandingly interesting -- selection force towards increased evolvability \cite{Stu37,Lei70,Joh99,AG06,WGK+09}. Selection for large mutation rates is essentially indirect (it is not adaptive \emph{per se}) \cite{Web96,SGJ+00}, and strongly depends on the recombination rate between mutator alleles and potentially-beneficial mutations: when such a positive-effect mutation occurs, its frequency will increase in the population, hence hitch-hiking the genetically-linked mutator. This mechanism is effective in low-recombination asexual organisms \cite{TTL+99,Orr00}, whereas recurrent crossovers makes adaptive evolution of high mutation rates unlikely in sexual organisms \cite{Kim67,Lei73,BMD07,Lyn08}, although the equilibrium rate might be non-zero \cite{IMI+89,Joh99b}. 

Yet, mutation rates in sexual multicellular eukaryotes tend to be higher than in asexuals, suggesting that genetic drift, genetic constraints, and external factors, rather than adaptation, are the main drivers of evolution of mutational evolvability in large organisms \cite{Lyn08}. In particular, a substantial part of genomic mutations can be attributed to the presence and the activity of Transposable Elements (TEs), which are self-replicating DNA sequences that can be found in the genome of virutally all organisms. Initially described as adaptive "controlling elements" (see \cite{Mcc84} for a full historical record), the deleterious consequences of TEs lead population geneticists to explain their ubiquity by their "selfish DNA" properties \cite{OC80,DS80}. Theoretical models confirmed that thanks to their intrinsic amplification ability, TEs are efficient genomic parasites of sexually-reprodcing organisms, able to invade populations and genomes in spite of an average deleterious effect \cite{Hic82, CC83, LC05}. 

TEs generate a broad spectrum of mutations, including gene disruptions, changes in regulation or alternative splicing patterns, recombinations, inversions, and segmental duplications or deletions. Most of these mutations are expected to be deleterious or neutral, but a few of them are likely to change the phenotype in an adaptive way, and be fixed by natural selection. Although this does not contradict the selfish DNA hypothesis as the primordial explanation for TE universality, the accumulation of empirical evidence for "domesticated" TE-related copies \cite{MMN+99,SII09} has generated a great deal of confusion about the cause-consequence relationship between TE activity and the associated increase in evolvability \cite{CGB+00}. As a matter of fact, little is known about the quantitative and qualitative consequences of introducing the possibility of beneficial mutations in TE theoretical dynamic models. 

In asexuals, TEs cannot spread as selfish DNA sequences, and theory confirms that their potential mutator effect can explain their presence \cite{SH86,MK97,MB02,EB03,MB06,SLC+13}, along with recurrent horizontal transfers \cite{Moo88,BM91,BBW10}. In sexual organisms, unconditional beneficial insertions seems to extend the time during which a TE family can maintain in a genome, but these copies are quicky inactivated by mutations and thus no longer participate to the TE dynamics \cite{LBC07,BLC12}. Virtually nothing is known about how such TEs affect the evolvability of the host species. 

With this paper, we aim at exploring the impact of transposable elements as mutators in sexual populations in constant and variable environments, using an explicit phenotype-fitness landscape based on the Fisher geometric model \cite{Fis30} with a moving optimum. 

\section{Methods}
\label{ch:methods}

\begin{figure}[hbt]

\begin{tabular}{cc}
\includegraphics[width=0.475\linewidth]{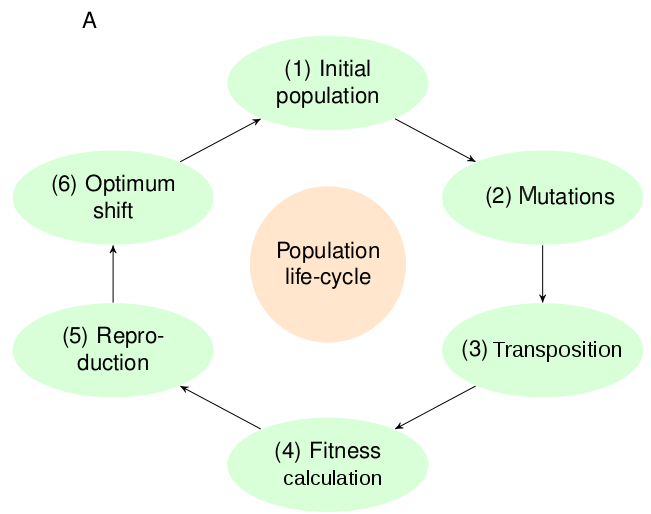} &  \includegraphics[width=0.475\linewidth]{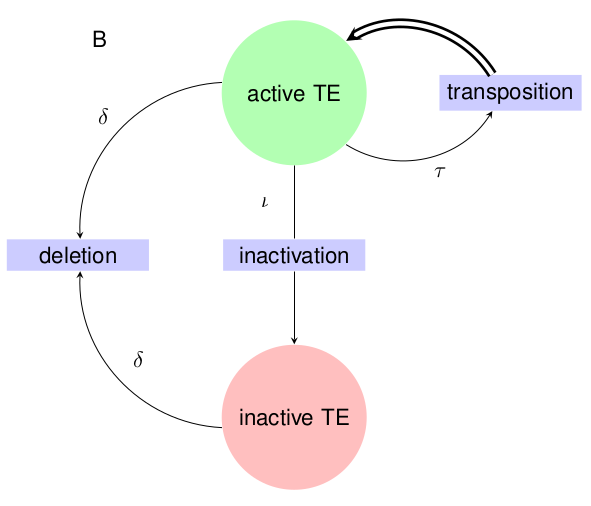}  
\end{tabular}
\caption{\scriptsize General overview of the stochastic model. (A) Presents main stages of a single generation's life-cycle: (1) initial population is composed of $N$ organisms, (2) the random, TE unrelated mutations modify base phenotype of each individual, (3) TE proliferate and affect TE-related phenotype, (4) summary of individuals' and population's fitness, (5) sexual reproduction resulting in a new generation, (6) shift of the optimum phenotype. (B) TEs proliferation events and their probabilities. Active copies can transpose resulting in new active copy or inactivate. Both active and inactive copies can undergo permanent deletion.}
\label{fig:model}
\end{figure}

We introduce a stochastic computational model that investigates the evolutionary dynamics of diploid populations of size $N$. Our main interest is focused on TEs activity, their influence on organism's phenotype and as a result on its fitness and adaptive abilities. Figure \ref{fig:model}A summarizes main stages that a population will follow in each generation. 

\subsection{Genotype - phenotype model}

Our framework is based on the Fisher's Geometric Model (FGM)
\cite{Fis30} with moving optimum \cite{KH09b}.
Each organism is represented by a phenotype $\phe$, which is 
represented as a real-valued, $n$-dimensional vector describing 
$n$ independent phenotypic traits. 
In accordance with the FGM, an organism corresponds to a point in $n$-dimensional
phenotypic space and each mutation changes its position by a $n$-dimensional vector drawn from the $n$-dimensional, multivariate centered normal distribution $\mathcal{N}(\bm{0}, \bm{I})$. Here, $\bm{I}$ is an identity matrix of size $n$, which expresses the assumption that mutations are fully isotropic and also determines a \textit{yardstick} in our model, as the \textit{variance of mutational effects}, $\sigma^2_m \equiv 1$.
The phenotype itself has two components: 
(1) the \textit{base phenotype} $\phe_b$ which is modeled as diploid, one-locus genome with real-valued alleles. $\phe_b$ changes only due to genome (TE-unrelated) mutations; 
(2) the \textit{TE-related phenotype} $\phe_T = \sum_{t \in T} \phe_t$, 
which is a sum of the mutational effects of all TEs of an organism.
As a result phenotype $\phe$ can be naturally expressed as $\phe = \phe_b + \phe_T$.
Both TE-related and TE-unrelated mutations are drawn from the same distribution. TE-unrelated mutations occur with a mutation rate $\mu$ (genomic mutation rate per generation), while TE-related mutations are directly related to TE activity (see below). 

\subsection{TEs dynamics}
For each simulation, the initial population
is composed of $N$ organisms hosting $10$ homozogous active TE copies. 
Each generation, TEs can undergo specific
events (also see Figure \ref{fig:model}B): 
(1) with a probability $\trate$ an active copy may transpose in a
\textit{copy-paste} manner, which results in an insertion of a new 
TE copy at some locus in a genome with a random mutational effect $\phe_t$;
(2) with a probability $\nu$ an active copy may become inactive and lose its
ability to transpose, but its mutational effect does not change;
(3) with a probability $\delta$ (deletion rate) both active and inactive copies can get
deleted from the genome. Along with this event the mutational effect is
cancelled.  


\subsection{Fitness and sexual reproduction}

We assume that the fitness function is multivariate normal and isotropic: 
\[
 \fit_{\theta}(\phe) = 
 \exp\left( 
 -\frac{ \vert\vert \phe - \phe_\theta \vert\vert^2 }{2 \sigma^2} 
 \right)
\]
where $\fit_\theta(\phe)$ is the relative fitness of an individual of (multivariate) phenotype $\phe$, and $\vert\vert \phe - \phe_\theta \vert\vert$ is the Euclidian distance between the phenotype $\phe$ and the optimal phenotype $\phe_\theta$. The variance of this Gaussian $\sigma^2$ is inversely proportional to the strengh of stabilizing selection (selection disappears when  $\sigma^2 \rightarrow \infty$). 

Sexual reproduction consists of picking two parents with a probability propotional to their fitness.  Each organism creates a gamete, 
which contains: one allele from the genome and set of TEs that will be inherited. Gamete composition follows the Mendel's inheritance rules and assumes genetic independence between loci. A new organism is created by merging both gametes. Its \textit{TE-related-phenotype} $\phe_T$ is calculated by summing up the effects of all inherited TEs, and the
\textit{base phenotype} is equal to:
\[
\phe_b = \frac{\phe^{(1)}_b + \phe^{(2)}_b}{2} + \varsigma_g
\]
where $\varsigma_g$ is a random deviate related to the genetic segregation variance \cite{Bul85}, distributed as $\varsigma_g \sim \mathcal{N}(0, s^2_g/2)$, where $s^2_g$ is the genetic variance in the parental population. We assumed that generations were not overlapping, and that the population size ($N$) was constant (soft selection). 

\subsection{Moving optimum}
Environmental change was modelled as a slow, gradual shift in the phenotypic optimum: 
$\phe_\theta (t) = t \cdot \kappa$, where $\kappa$ describes 
the speed of the environmental change. When $\kappa = 0$, the optimum remains at its starting position (arbitrarily set at $\phe_\theta (0) = 0$), which corresponds to stabilizing selection. 

\subsection{Equilibrium state}
In order to determine if the population is in the equilibrium state with
respect to the TE copy number, we compute the slope of the regression line
(i.e. we build a linear model $Y = \beta X$ predicting TE copy number from the last  $G_{reg}$ generations, basing on the generation number). 
We assume that equilibrium state is attained when the regression coefficient satisfies $\vert\beta\vert < 0.02 $.

\subsection{Simulations setting}
\begin{table}

\caption{Model parameters used for simulations. }

\begin{tabular}{L{3cm}C{0.75cm}C{1.25cm}C{1.25cm}C{1.25cm}C{1.25cm}}
\hline
Parameter name & Notation & Selfish (S) & Non-Selfish (NS) & Selfish with Inactive (SI) & Non-Selfish with Inacitve (NSI) \\
\hline
\multicolumn{6}{c}{Environmental parameters} \\
Population size & $N$ & \multicolumn{4}{c}{2000} \\
Number of generations & $G$ & \multicolumn{4}{c}{10000} \\
Number of phenotypic traits & $n$ & \multicolumn{4}{c}{2} \\
Selection strength$^a$ & $\sigma$ & \multicolumn{4}{c}{$10$} \\
Optimum change$^a$ & $\kappa$ & \multicolumn{4}{c}{ stab. sel. = $0$ / moving opt. = $0.002$} \\

\multicolumn{6}{c}{Genomic parameters} \\
Background mutation rate$^b$ & $\mu$ & \multicolumn{4}{c}{0.003} \\
Transposition rate$^c$ & $\trate$ & \multicolumn{4}{c}{0.002} \\
Deletion rate$^c$ & $\delta$ & 0.0002 & 0.002 & 0.0002 & 0.002 \\
Inactivation rate$^c$ & $\iota$ & 0 & 0 & 0.002 & 0.002 \\
\hline
\end{tabular}
{\scriptsize \\$^a$ some parameter values were chosen to ensure that simulations converge to an equilibrium state; $^b$ per genome and per generation; $^c$ per copy and per generation.}

\label{tab:parameters}
\end{table}

Individual-based simulations were performed in order to understand the complex behaviour of our model. Outcomes presenting individual results along with significant statistics were collected and can be visualized at \url{http://bioputer.mimuw.edu.pl/TE-model}.
Additionally, the setting for main parameters that are investigated throughout the next section are presented and described in Table~\ref{tab:parameters}.
Finally, in the next section we present exploration of the parameter space for different scenarios which was performed using the Monte Carlo sampling.

\section{Results}
\label{ch:results}
Our main purpose is to understand the complex relationship between genotype, phenotype and fitness in the context of TE proliferation, within the framework of the FGM.
We aim at describing the evolution of TE copy number along with fitness in sexual populations, especially when taking into account environmental change. 

\subsection{TE proliferation dynamics}

\begin{figure}[thb]
\centering
\begin{tabular}{c}
\includegraphics[width=1
\textwidth]{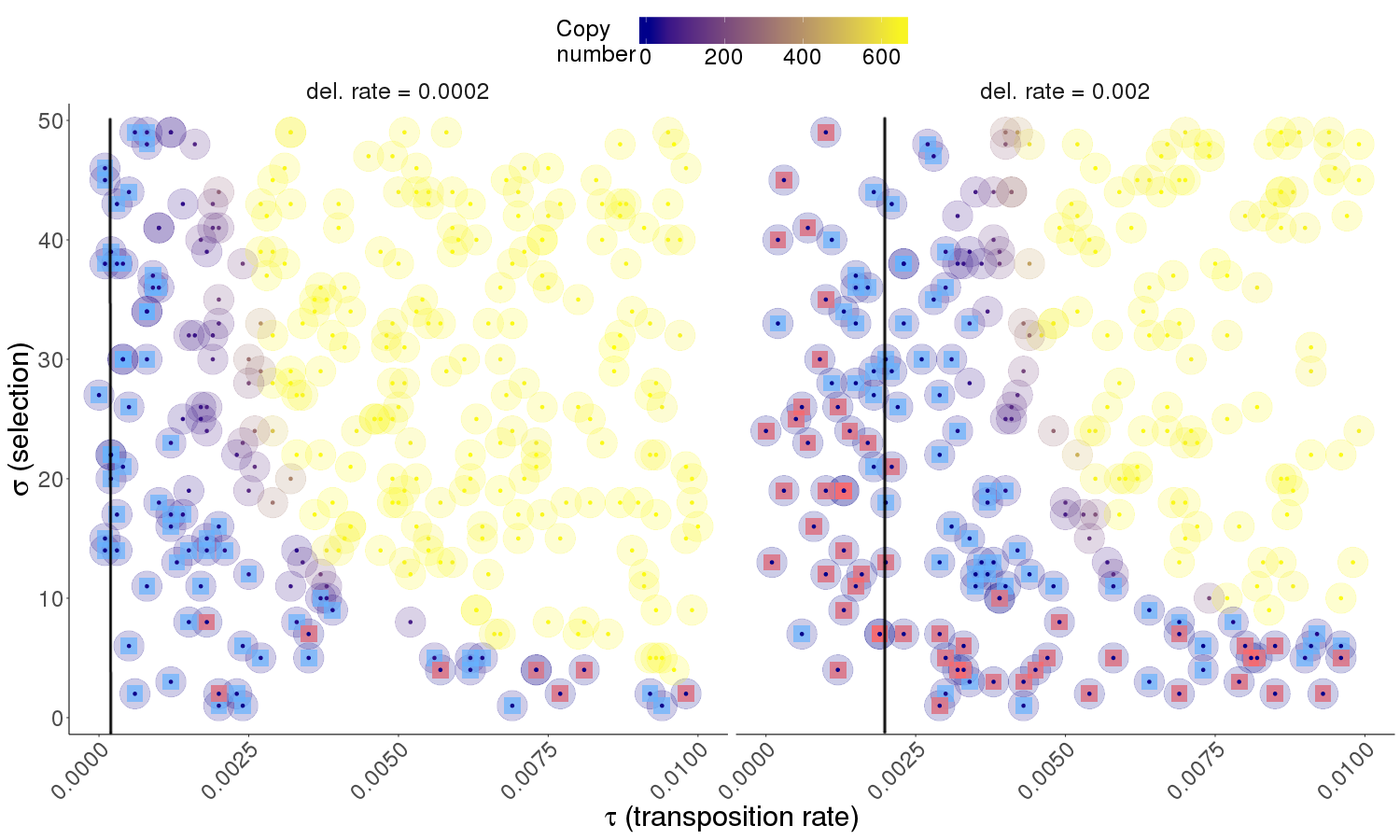} \\

\centering
\end{tabular}
\caption{ \scriptsize 
TE dynamics in stabilizing selection model for two different deletion rate levels ($\delta$). Left and right panel represent, $2\cdot 10^{-4}$ and $2\cdot 10^{-3}$, respectively. Additionally, black lines point out these values on x-axis. 
Continuous color scale encodes the mean copy number in population at the end of simulation (10,000 generations). Scenarios for which the non-zero equilibrium is attained are depicted by blue squares while trivial equilibria (for zero TE copy number) red.\\
Note, selection gets weaker with increasing $\sigma$.
%
}
\label{fig:stab_sel_tratevs_sel}
\end{figure}

The exploration of the parameter space was performed for both the stabilizing
selection (i.e. the optimal phenotype in time is constant, $\kappa = 0$), 
and the moving optimum scenarios. At the very begining, the value of all parameters except
the transposition rate ($\trate$) and the selection strength ($\sigma$) were fixed (see Table~\ref{tab:parameters}) in order to determine ranges for $\trate$ and $\sigma$ in which a \textit{transposition-selection equilibrium} is attainable. 
The monte Carlo sampling, yielded three regions corresponding to different model behaviours (see Figure~\ref{fig:stab_sel_tratevs_sel}).

\begin{figure}[thb]
\centering
\begin{tabular}{c}
\includegraphics[width=1\textwidth]{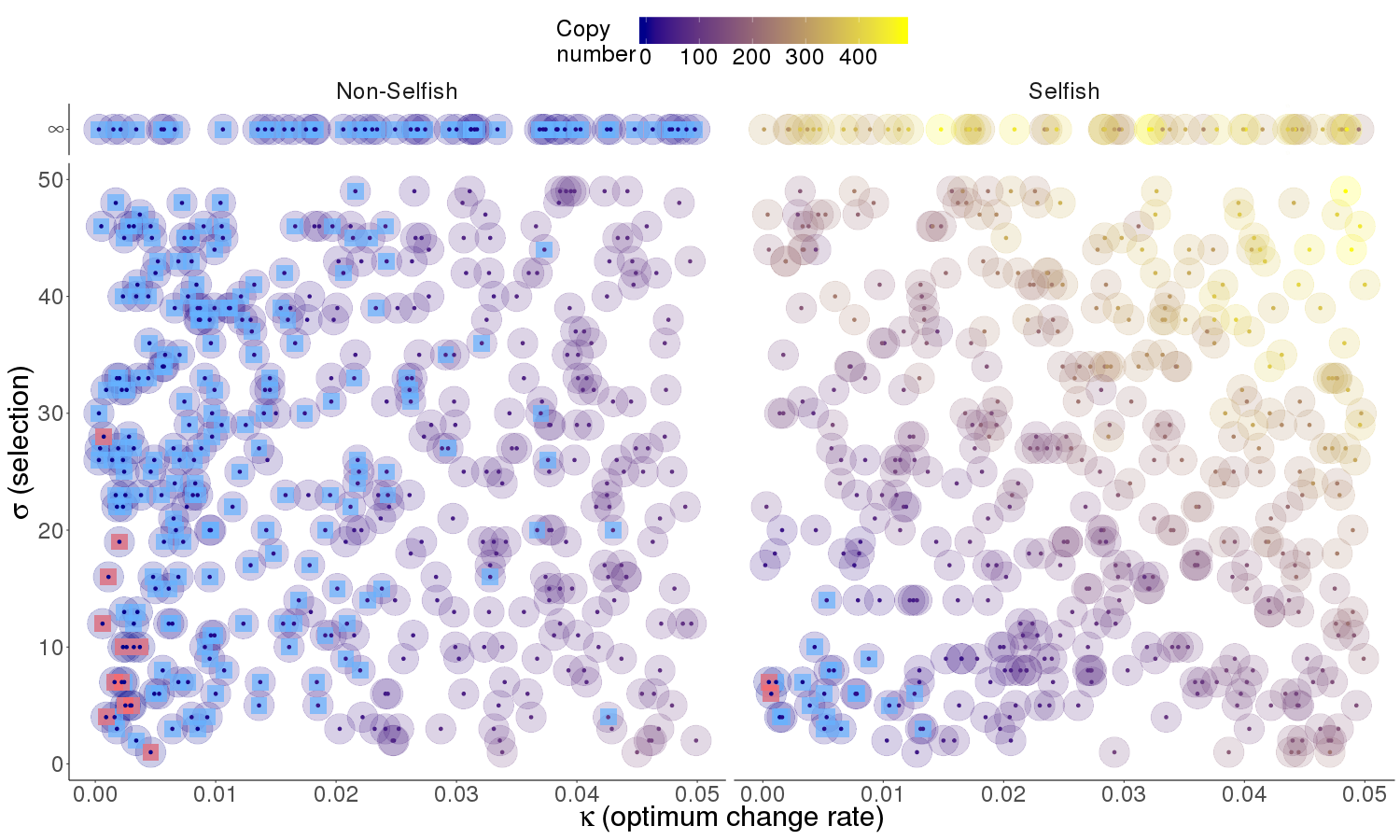} \\

\centering
\end{tabular}
\caption{ \scriptsize 
Comparison of TE dynamics for two scenarios in the moving optimum model (\textit{non-selfish} vs \textit{selfish} on the left and right handside, respectively). 
A color of each circle encodes the mean TE copy number in population at the end of simulation (10,000 generations) for given parameters $(\kappa, \sigma)$. Scenarios for which the non-zero equilibrium is attained are depicted by blue squares while trivial equilibria (for zero TE copy number) red.
Note, selection gets weaker with increasing $\sigma$.
%
}
\label{fig:moving_opt_change_vs_sel}
\end{figure}

The TE dynamics has been classified into three cases according to the variability in the last $200$ generations out of $10000$ for each simulation: (i) uncontrolled increase (i.e. exponential growth),   (ii) equilibrium in TE copy number, (iii) loss of all TEs. 
  
Interestingly, the subset of parameters for which the model upholds a \textit{transposition-selection equilibrium} is quite narrow . Very strong selection ($\sigma < 5$) reinforces the occurrence of TE maintenance in populations. Indeed, selection starts to promote the best reproducers very fast, which leads to homogenous population of homozygous organisms with fixed copy number. Loose selection ($\sigma > 15$) also allows easy maintenance of TEs, because the selective cost of TEs is moderate, preventing their elimination by natural selection. As the transposition-selection equilibrium has been of historical interest, especially in the early models, we picked for further investigation $\trate=0.002$ and an intermediate value of $\sigma = 10$ that lead to a stable polymorphic copy number under stabilizing selection. 

An analogous Monte Carlo sampling procedure has been performed for the moving optimum model. The $\trate$ parameter was fixed and we explored the subspace defined by $\sigma$ and $\kappa$  for meaningful ranges. These simulations were run with both \textit{non-selfish} and \textit{selfish} TEs, i.e. $\delta = \tau$ and $\delta = \frac{1}{10} \tau$, respectively. Figure~\ref{fig:moving_opt_change_vs_sel} presents the different TE proliferation dynamics observed in the full parameter space along with mean fitness distribution in populations in question.


 

As the strength of natural selection against deleterious and beneficial TE insertions is the main driver of TE dynamics in our model, we investigated two important parameters influencing the probability of allele fixation: (i) the number of traits, which conditions the balance between beneficial and deleterious mutations, and (ii) the population size, which affects the probability of a mutation to be effectively neutral. 

\begin{figure}
\centering
\begin{tabular}{c}
\includegraphics[width=1\textwidth]{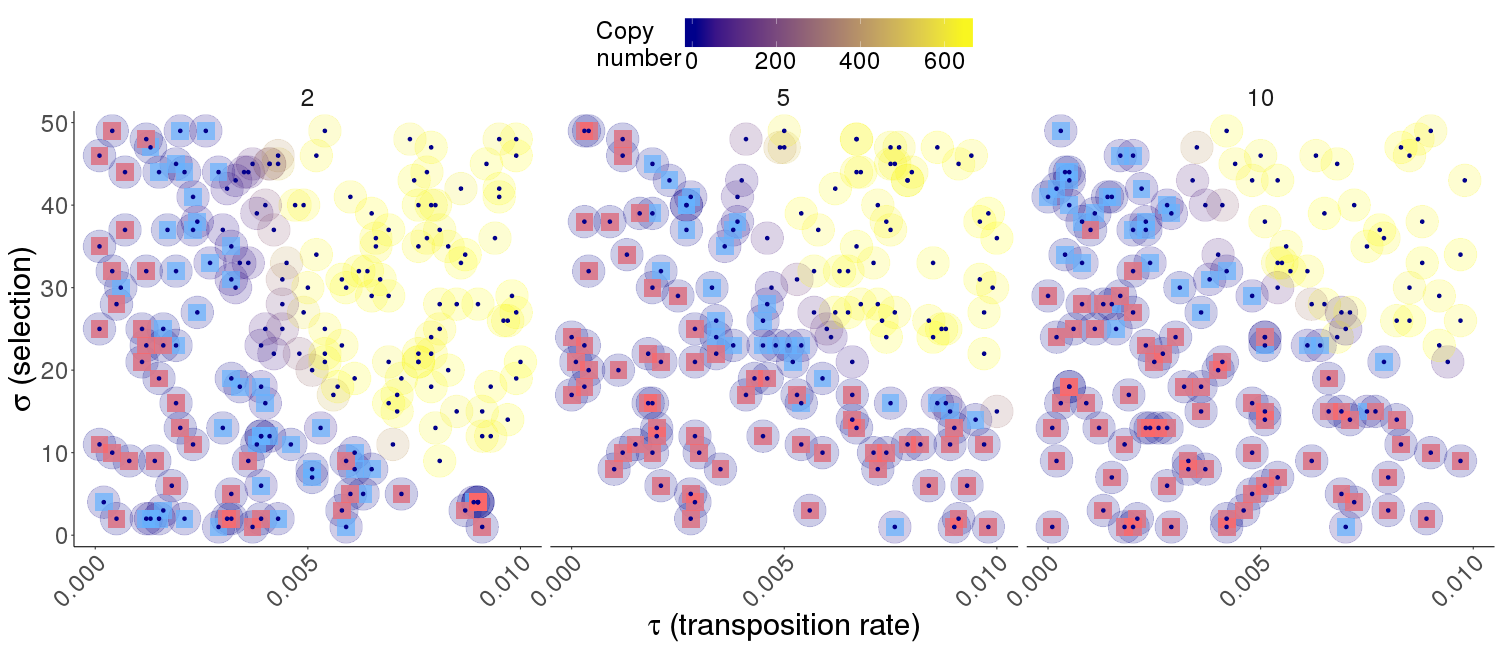} \\
\centering
\end{tabular}
\caption{ \scriptsize TE copy number at the end of simulations as a function of organism's complexity ($n$) assuming \textit{stabilizing-selection} scenario. Above figures present the impact of increasing the number of phenotypic traits of an individual on TE dynamics (upper panel). A color of each circle encodes the mean TE copy number in population for given parameters $(\kappa, \sigma)$. Scenarios for which the non-zero equilibrium is attained are depicted by blue squares while trivial equilibria (for zero TE copy number) red ones. One can notice that the equilibrium states require higher values of transposition rate and less rigorous selection when $n$ increases (note, selection gets weaker with increasing $\sigma$). }
\label{fig:traits}
\end{figure}

As mutations in the FGM affect all traits in an isotropic way, the probability that a specific mutation decrease the Euclidian distance to the phenotypic optimum decreases with the dimensionality of the phenotypic space~\cite{curseofdim}. Increasing the number of phenotypic traits in the simulation is consistent with this theoretical expectation, as the transposition rate allowing TE maintenance increases accordingly (Figure~\ref{fig:traits}). Further simulations were run with two phenotypic traits, the first trait being affected by the  environmental change, while the second trait optimum remained constant. 

\subsection{Active and inactive copies}

Most copies in eukaryotic genomes are inactive (unable to transpose). Such copies do not impact the transposition dynamics directly, but their impact of the phenotype remains. Inactivation of the transposition ability of TE copies is thus likely to be an important phenomenon, as it cancels the mutagenic activity of beneficial TE insertions. 

Inactive copies were introduced to the system by setting a mutation rate from active to inactive TE copies ($\iota > 0$). Two different deletion rates were considered ($\delta_{S} = \frac{1}{10} \cdot \tau$ and $\delta_{NS} = \tau$). This settels four scenarios: (i) $S$ -- selfish, transposition rate is one order of magnitude larger than deletion rate; (ii) $NS$ -- non-selfish, transposition rate and deletion rate are of the same order (here, equal); (iii) $SI$ -- selfish with possibility of transposon inactivation; (iv) $NSI$ -- non-selfish with possibility of transposon inactivation.
 
\begin{figure}[bht]

\centering
\begin{tabular}{ccc}
\includegraphics[width=0.3\textwidth]{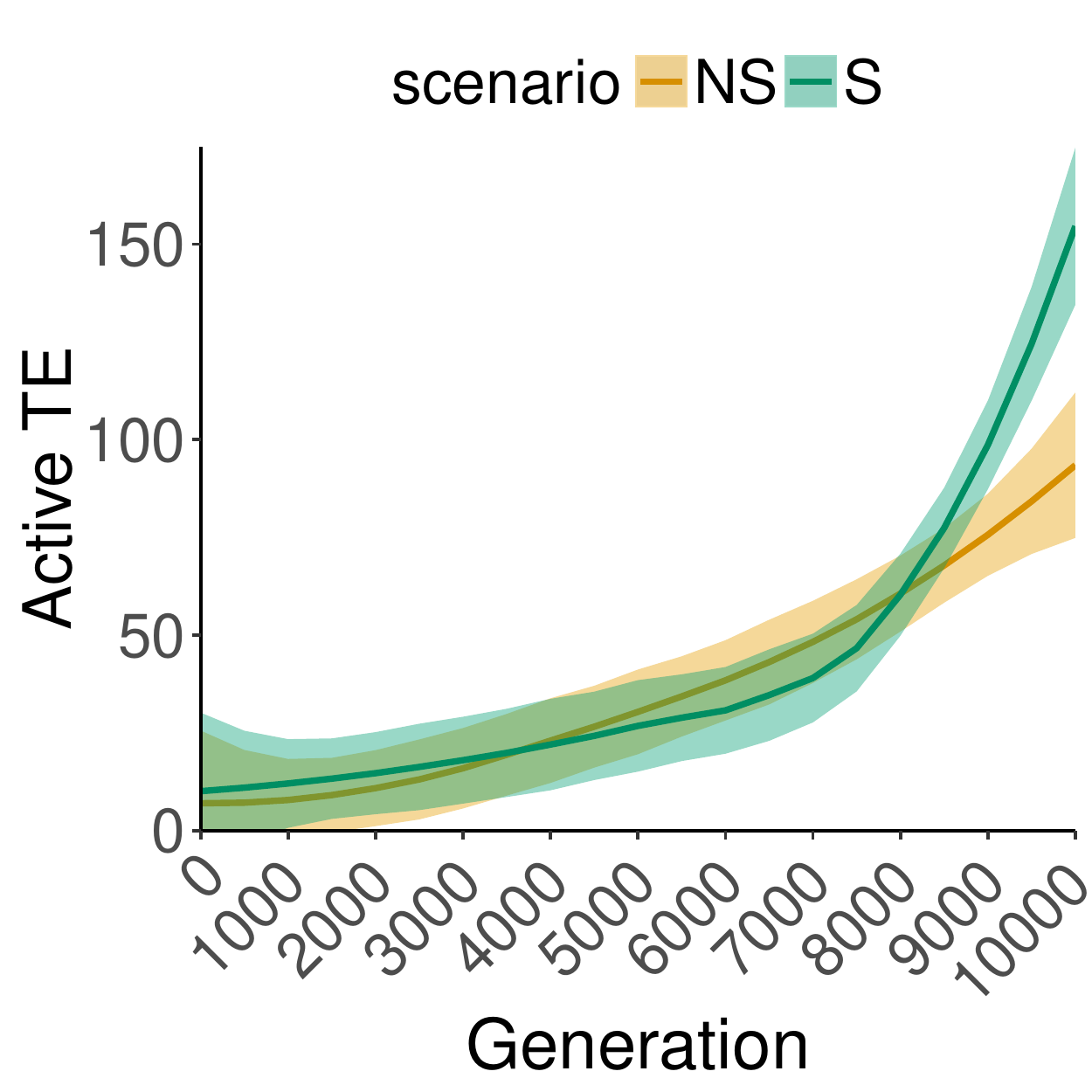} &
\includegraphics[width=0.3\textwidth]{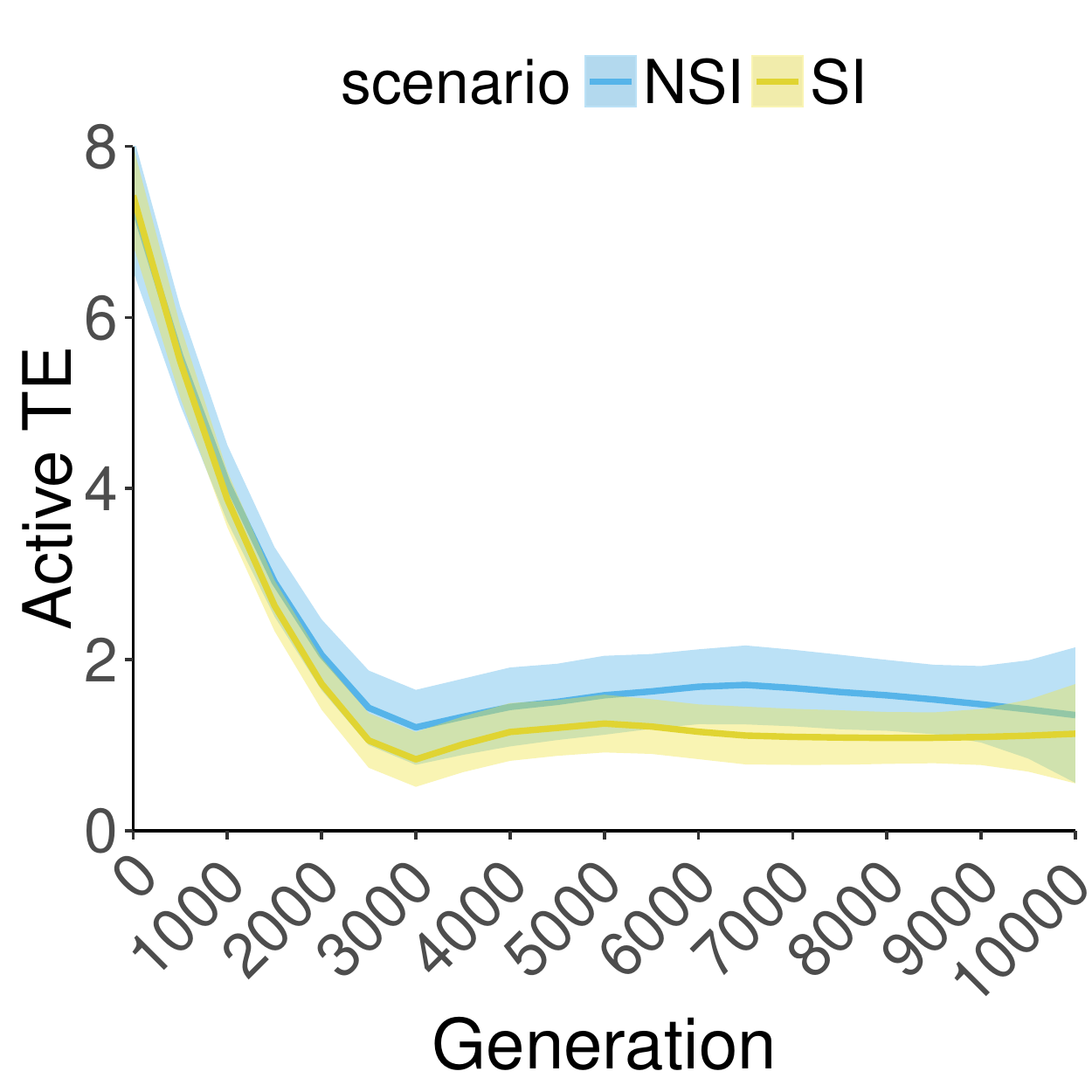} &
\includegraphics[width=0.3\textwidth]{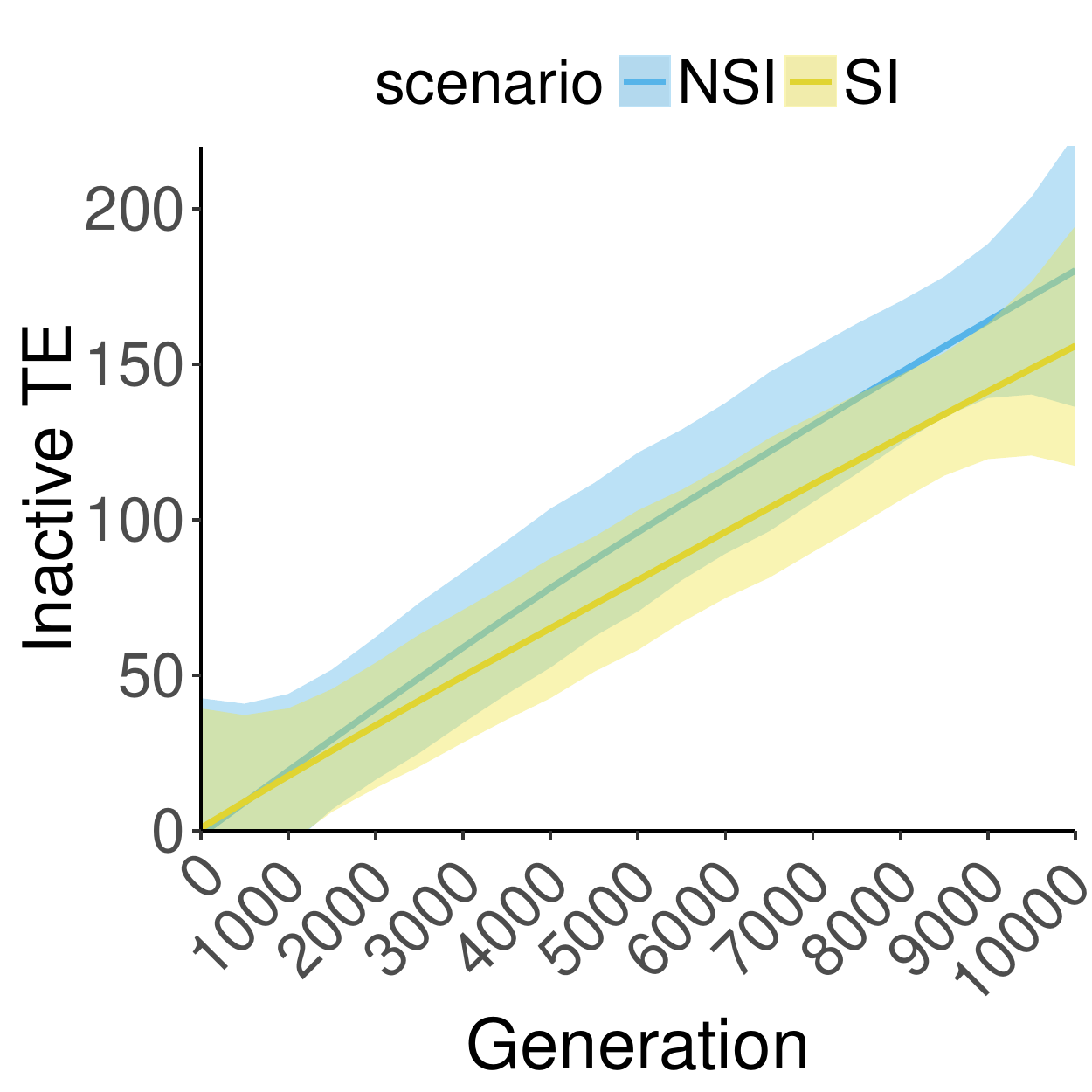}
\centering
\end{tabular}
\caption{Trend in mean copy number of both active and inactive TE copies throughout the whole simulation. The first plot (from the left) presents the trend of active copies in selfish and non-selfish scenarios. The second one also presents the trend of active copies providing the information that populations in which TEs can inactivate carry only few copies of active TEs (see the y-axis values). 
Finally, the third plot provides the information about linear accumulation of inactive copies. 
Each line in plot is a smoothed conditional mean basing on \textit{geom\_smooth} from R \textit{ggplot2} package, that is why on the first plot lines do not start at the same level even though for each simulation active TE copy number is equal to 10. The shade around the each line displays the confidence interval for $0.95$ level. Setting: $\sigma = 10.0$, $\tau=0.002$, $\iota = 0.003$}
\label{fig:teintime2}
\end{figure}

With inactive copies appear a new type of dynamics featuring a linear increase, which is a peculiar steady-state equilibrium. We know from~\cite{cmc91} that neutral active copies cannot accumulate without increasing the genetic load, but inactive copies can. As one can observe on Figure~\ref{fig:teintime2}, the introduction of inactive copies pushes active ones to the background and their number is usually on a very low but stable level, while inactive ones are slowly aggregating in a linear fashion. 

\subsection{Non-linear models for average fitness}

Our next goal was to investigate the complex link between adaptation, mutation rate, and the evolution of evolvability. One of the main differences between mutation rate and TE-driven mutations is the fact that the amplification ability of TEs (as selfish DNA) generates a natural trend towards an increase of the mutation rate, which is independent from adaptation. In this context, we focused on how model parameters (TE copy number, strength of selection and transposition rate (\textit{stabilizing  selection} scenario)/speed of environmental change (\textit{moving optimum} scenario) affect the mean fitness of the population at equilibrium.

The mean fitness of the population has two major components: (i) the average distance of the population from the optimum (how well the population is adapted to its current environment), and (ii) how spread are the individuals in the population (featuring the population evolvability). Interestingly, in our moving optimum setting, these two properties merge as populations have reached a steady state (i.e. how well they are adapted to the current optimum measures the same thing as how well they will adapt to a new environment in the next generation). We thus propose to use the average fitness at steady state as a measurement of both adaptation and evolvability. 

For stabilizing selection model, we consider the mean fitness as an unknown function of TE copy number, selection strength and transposition rate. In case of moving optimum scenario the transposition rate parameter is replaced by the speed of environmental change.
Since assuming that all predictors contribute to the response variable (i.e. average fitness)  only in linear and additive manner is non-realistic (see $\fit$ definition) we decided to fit non-linear regression model as implemented in the Random Forest approach see~\citep{Breiman:2001}

\begin{figure}[thb]
 \centering
 \begin{tabular}{c}
 \includegraphics[width=1\textwidth]{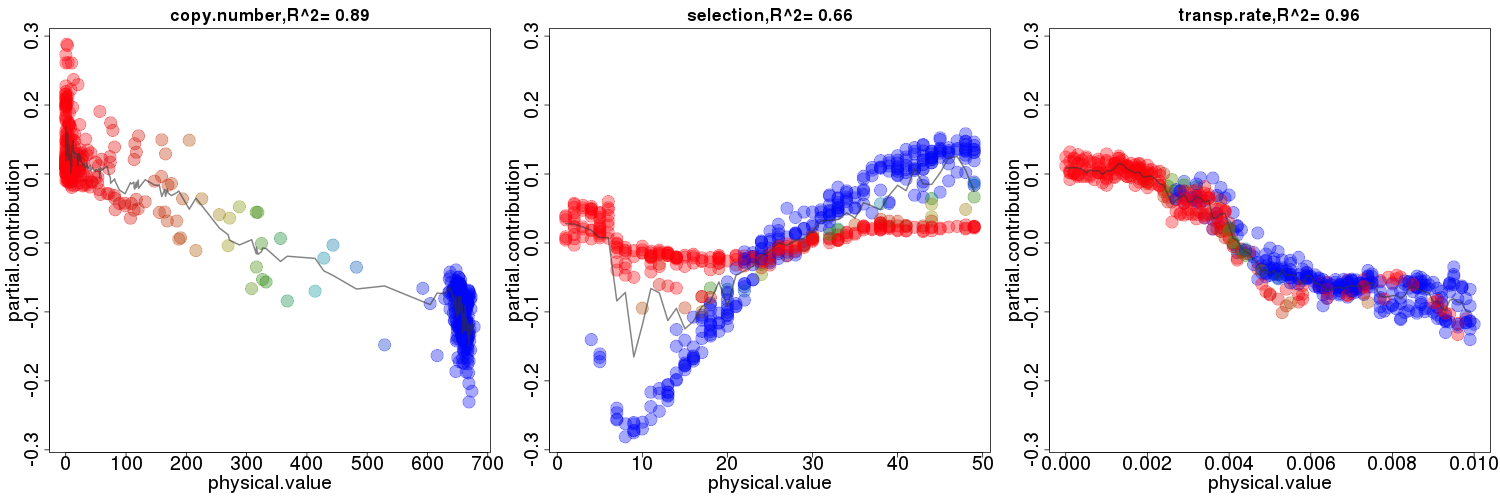} \\
 \includegraphics[width=1\textwidth]{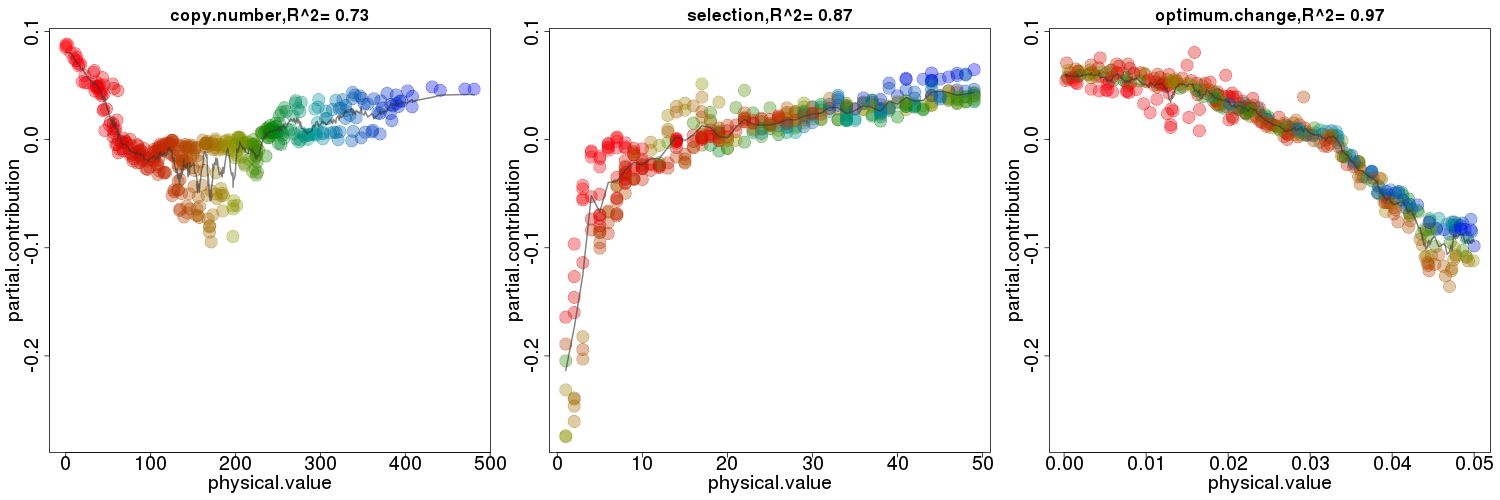} \\
 \includegraphics[width=1\textwidth]{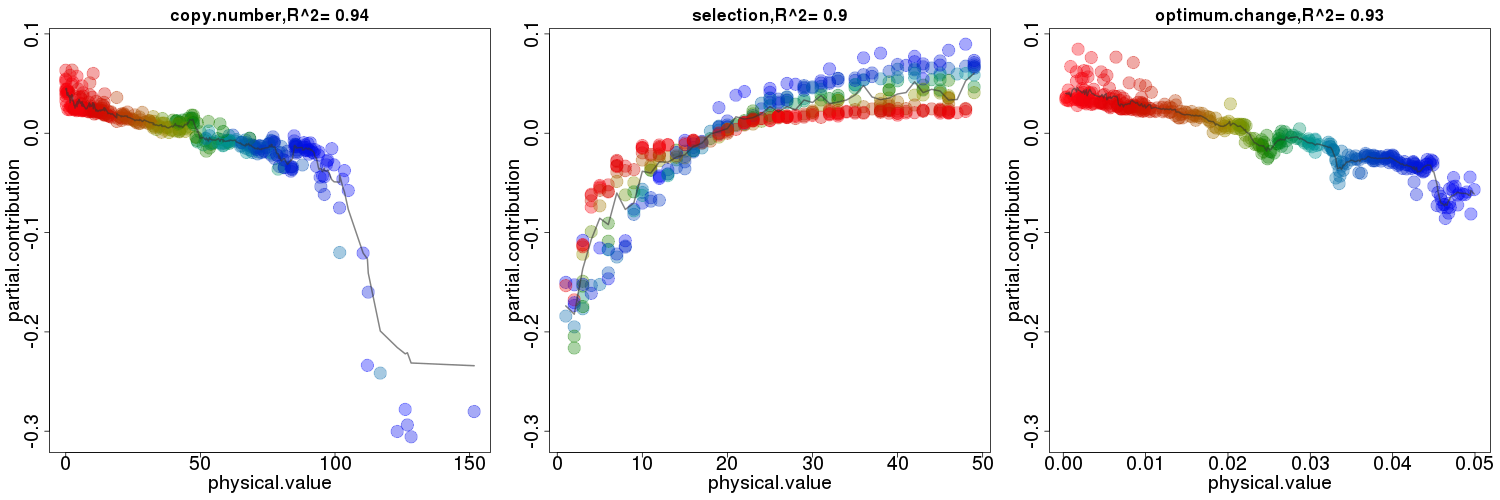}
 \centering
 \end{tabular}
 
 \caption{Influence of various variables on mean fitness (measured  as the change of fitness predicted, due to splits in Random Forest by a given variable). From top: stabilizing selection, moving optimum \textit{selfish} scenario, moving optimum \textit{non-selfish}
scenario. The color gradient corresponds to TE copy number.}
 \label{fit:nlmfitness}
\end{figure}

Figure~\ref{fit:nlmfitness} gives insights into  the nature of nonlinear contribution for all considered predictors. Each point corresponds to one simulation obtained in Monte Carlo uniform sampling scheme. The x-axis spans the range of given predictor's values and y-axis coordinate describes the
cross validated predictor contribution (change of predicted value due to predictor value, i.e. the influence to the mean fitness). 

The color gradient for each figure encodes the change of TE copy number. It turns out that for stabilizing selection (three panels in upper row on Figure~\ref{fit:nlmfitness}), this predictor covers the most of the data variation independently identified by principal component analysis (PCA). 
For readability we keep this color-code also in moving optimum scenarios. Summarizing, the analyzed data are the same as
the dataset analyzed in Figure~\ref{fig:stab_sel_tratevs_sel} and~\ref{fig:moving_opt_change_vs_sel}, and the regression analysis give insights   how the given predictor 
(one panel) affects the mean equilibrium fitness. 

The color of data points changes gradually with increased TE copy number (three panels in the left column of Figure~\ref{fit:nlmfitness}). 
For stabilizing selection and non-selfish moving optimum scenario the influence of TE copy number plot has same monotonically decreasing shape, 
i.e. moderate number of TE copies are beneficial, while more copies tend to decrease the average equilibrium fitness. Slightly different plot is obtained for selfish scenario 
(middle panel): 
the parabolic shape suggests the positive contribution of this parameter in relatively low and high TE copy number, while the moderate TE copy number has negative impact on the equilibrium fitness. 

The analogous monotonically decreasing behavior can be observed for transposition
rate and speed of optimum change parameters (three panels in the
right column in Figure~\ref{fit:nlmfitness}). 
Over the range of variation of the parameters, high
transposition rates (and high environmental stress) are deleterious.

The influence of selection parameter $\sigma$ is more intriguing. The interesting
trend can be observed for stabilizing selection (top middle panel in Figure~\ref{fit:nlmfitness}).
Two subpopulation can be distinguished: for simulations with low TE copy
number (red points) the selection parameter does not influence the mean fitness,
while for simulations carrying more TE copies (blue points) the selection
parameter strongly affects the response variable (loose selection is beneficial).
For moving optimum scenario the selection parameter influences fitness in
monotonically increased manner (i.e. more loose selection is more beneficial).



%

\subsection{Adaptation and Transposable Elements}


%
%

Next, we studied how in general TEs influence evolvability. We observed a tradeoff resulting from mutagenic activity. On one hand, populations deprived of TEs are homogenous, and polymorphism is only driven by TE-unrelated mutations (with rate $\mu$). On the other hand, in presence of TEs, populations display a higher mutation rate, wider phenotypic ranges, and thus a larger evolutionary potential. Depending on the parameters and the number of TEs at equilibrium, populations can display lower or higher average fitness than in the non-TE case. However, when the TE-unrelated mutation rate is too low to allow the population to track the optimum efficiently, the presence of active TEs may substantially improve the average equilibrium fitness.


\begin{figure}[ht]
\includegraphics[width=1\textwidth]{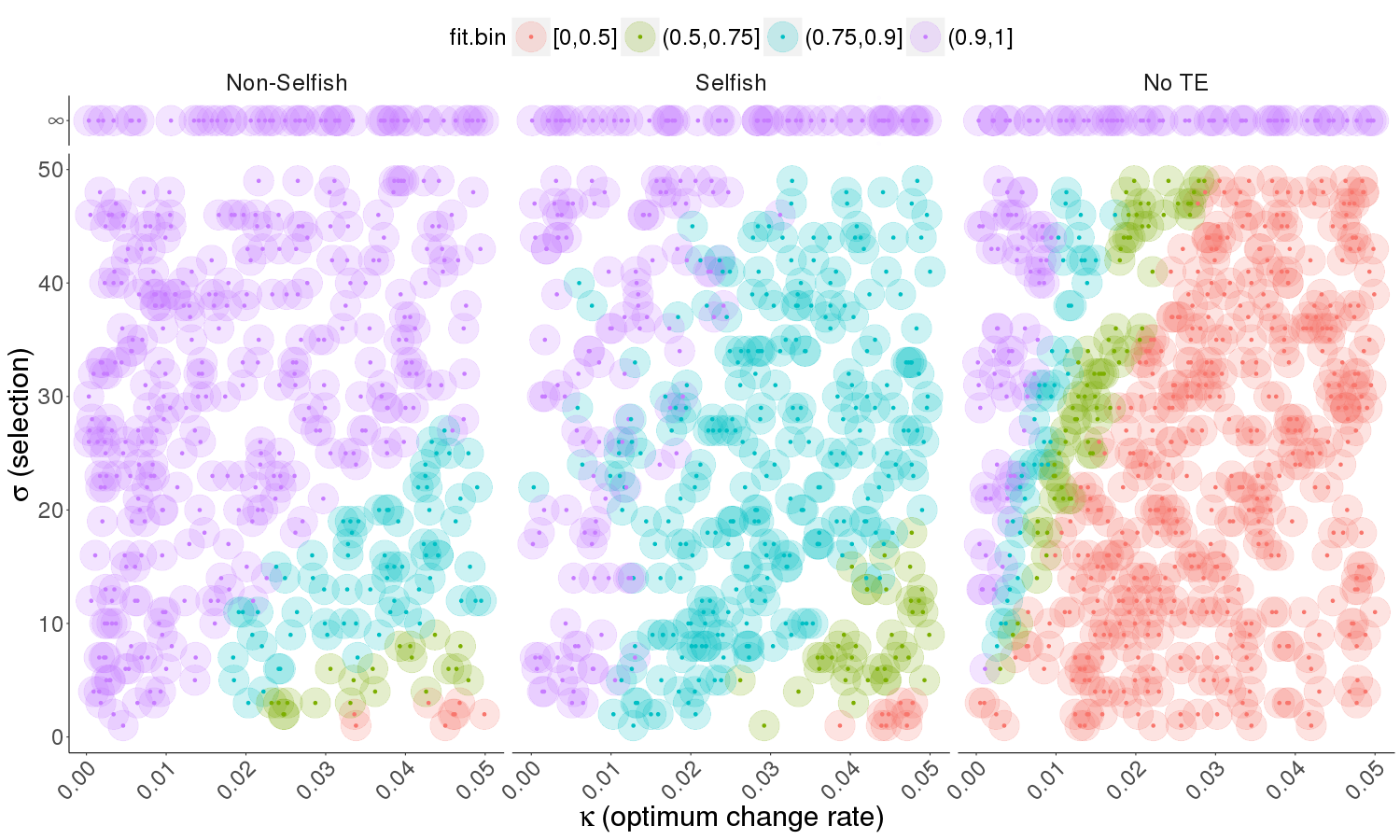} 

\caption{The average fitness in populations with different TE dynamics. 
Color of each circle, in each plot describes the average fitness in population at the end of simulation. Each plot itself correspons to specific scenario (from the left: \textit{Non-selfish}, \textit{Selfish}, \textit{No TE}).
}
\label{fig:fitness_with_vs_without}
\end{figure}

%
%


We already know that the presence of TEs in population is on average beneficial when it comes to overcoming the environmental pressure. 
In the figure~\ref{fig:fitness_with_vs_without} we present the patterns of the adaptive abilities in the sense of mean fitness value along with its trend in populations with and without TEs for different values of $\sigma$ (selection strength) and $\kappa$ (speed of the optimum change). 

In sum, TEs can introduce additional phenotypic variation that allows population to track the phenotypic optimum, at the expense of an increase in the variance in fitness (and thus, a higher fraction of maladapted individuals). 

\section{Discussion}
\label{ch:discussion}
 
\subsection{Modeling assumptions}

One of the original points of our modeling approach is the use of the Fisher Geometric Model as a genotype-fitness model, which explicitly accounts for the complexity of the genotype-phenotype relationship combined with a fitness landscape. The major consequences of this modeling choice on the dynamics of TEs are that (i) the effect of a TE insertion on fitness is no longer fixed, as it depends on the genetic background of the individual and the fistance from the phenotypic optimum; (ii) the model accounts for biological complexity (epistasis for fitness, pleiotropy...); (iii) the balance between the need for evolvability and the cost of mutations is straightforward. In spite of being widely used in theoretical evolutionary biology \cite{Ten14}, the FGM has been seldom implemented to study TE dynamics \cite{SLC+13}. Historical models of TE evolution instead used to consider either neutral or equally-deleterious effects for all copies \cite{CC83,Cha91,LC05}, while the possibility for beneficial effects was modeled more recently \cite{LBC07}, motivated by empirical observations. Yet, such beneficial effects were considered to be unconditionally adaptive, and such mutations occurred at the same rate whatever the average fitness of the population. Applying the FGM to TE dynamics releases these unnecessary assumptions, the cost being the lack of theoretical reference (in terms of e.g. transposition-selection equilibrium) compared with the most traditional literature. 

One of the major theoretical contribution of the FGM is the detailed study of the adaptive process towards a fixed phenotypic optimum \cite{Orr98,Orr06}. The possibility that this optimum itself might change in the course of time changes drastically the dynamics of the population, as adaptation becomes a continuous process inspired by theoretical (e.g. the Red Queen process) or empirical (global warming) observations. Of course, such a never-ending directional shift in the optimal phenotype is unrealistic, but remains a good approximation of adaptation in a changing environment, provided that the direction of the optimum change does not fluctuate rapidly. We are thus confident that our simulation results catch some biologically-relevant mechanisms associated with long-term environmental modifications, such as climate change. 

In our model setting, phenotypic traits are expressed in mutational standard deviation units (i.e. the model is scaled by the mutational variance-covariance matrix). In this context, the isotropic fitness landscape has to be interpreted as a consequence of the fact that the mutational properties of traits are expected to evolve towards mutational covariances that match fitness covariances \cite{JBA14}. Model parameters need to be defined in terms of mutational steps; for instance, in the default parameter set, the rate of optimum change ($\kappa=0.002$) corresponds to around 4\% of the background mutational standard deviation ($= \sqrt{\mu \sigma^2_m}$). In other words, even a population deprived of TEs is likely to be able to track the optimum closely (provided that the number of traits $n$ is not too large), as the genetic diversity brought by mutations every generation is geometrically wider than the optimum change. In contrast, optimum changes of the order of $\kappa > 0.01$ constitute a challenge for TE-deprived populations.

Here, contrary to \cite{MK97}, TE-related and -unrelated mutational effects were drawn in the same distribution. Our purpose was to assess whether TEs could maintain in populations without providing them a special effect (such as the ability to cross fitness valleys). In practice, little is known about the relative fitness consequences of TEs vs.~ point mutations in genomes ; some huge molecular changes (such as inversions and translocations) could have minor or no impact of fitness, while single substitutions can be lethal. Nevertheless, the propensity of TEs to be a source of evolutionary innovation (such as for the immune system of vertebrates, \cite{AES98}) suggests that the nature of molecular changes might have very long-term consequences for genomes, albeit this level of evolvability remains a modeling challenge. 
 
\subsection{Transposable elements as evolvability helpers}

The presence and the activity of transposable elements in genomes promotes an increase in the mutation rate, which necessarily affects genome evolution. However, most mutations are deleterious, and individuals harboring a high mutation rate (due to e.g. a large amount of active TEs) will, in average, have a lower fitness relative to the rest of the population, leading to the elimination of mutagenic agents by natural selection. The need to adapt continuously to new environmental challenges may help "mutators" to invade asexual populations by hitch-hiking rare beneficial variants, but recombination break this association in sexual populations, making it possible to fix the advantageous allele while losing the deleterious mutator. According to this theoretical context, evolutionary helpers are expected to be rare or absent in sexual species. 

However, the properties of transposable elements makes them particularly interesing, since contrary to other mutators that increase the genomic mutation rate in general, TEs replicate in genomes, ensuring a complete linkage disequilibrium between the mutated and the mutator locus. Unfortunately for the "evolvability helper" hypothesis, every beneficial mutation also increases the mutation rate, inflating the genetic load up to the point that the beneficial effects vanish in front of the catastrophic consequences of an uncontrolled amplification of transposable elements. In our simulations, we found no evidence of parameter sets for which active TEs could be considered as adaptive in average, i.e. that TE copy number was positively correlated with relative fitness. In contrast, the accumulation of inactive adaptive copies was systematic when a non-nil mutation rate between active and inactive TEs was introduced -- in other words, when possible, evolution tends to decouple the mutator activity itself, which is deleterious, and the advantageous effect of specific insertions. 

Nonetheless, some simulations depicted interesting scenarios that could be compared to empirical knowledge about genome content. First, some simulations implementing a fast environmental change could display a temporary gain in average fitness when TEs were present. Although TEs remain maladaptive within populations, and thus do not invade populations due to positive Darwinian selection, there might be a transcient, but substantial, advantage for species carrying more TEs when adapting to a violent environmental change when competing with other species on the same resource. However, in addition to invoking species selection --- a mechanism which biological relevance is not firmly grounded, this hypothesis relies on the assumption that TE-unrelated mutations are limiting for evolution, which remains highly speculative. Second, the accumulation of beneficial, inactive copies (or, alternatively, active copies quickly inactivated by mutations) is realistic under a wide range of parameter values. In this case, active copies are maintained in low copy number due to their "selfish DNA" properties, but non-deleterious inactive copies (which are inert in the genome) can be maintained on the long term. This reminds the TE content of many eukaryotic genomes, in which potentially active copies are vastly outnumbered by TE relics. This also echoes the accumulating evidence for recurrent adaptive TE-derived mutations during evolution, such a molecular exaptation being the consequence (rather than the cause) of the universality of TE activity. 

It was postulated that TE activity might play a significant role in adaptive evolution~\cite{CG2013}. Founder populations colonizing novel ecological niches are usually of small size and reduced genetic variability, thus stress-induced TE activity may be a major factor required for rapid adaptation ~\cite{SSD2015}, directly corresponding to the results produced by our model. It could be one of possible explanations for the 'genetic paradox of invasive species', i.e. the potential for rapid adaptation despite the initial low genetic variability. Examples of the adaptive role of TEs in the process of colonization by Drosophila have been discussed recently~\cite{BFPG2014}. There is also a growing body of evidence in other organisms that increased copy numbers of certain TE families reflect their TE-aided adaptation to stress ~\cite{VALB1998, KTINS2000, NCYCY2006, SKEZA2014, ZZDSH2014}.

It is widely accepted that TE activity can be induced by a range of biotic and abiotic stresses, because of the relaxation of epigenetic control, which in the light of the results generated by our model, provides an additional regulatory level to evolvability. TE-mediated adaptation would therefore comprise the following stages: (1) stress-induced TE activation, (2) burst of activity resulting in copy number increase, (3) TE-induced beneficial rearrangments, and (4) TE inactivation. On the host side, it would correspond to (1) increase of the initial genetic variability, (2) natural selection for more fit individuals, (3) vertical transmission nad fixation of beneficial variants, (4) vertical transmission of other (neutral) TE copies resulting in higher copy number and, possibly, genome inflation. Even though only the rare beneficial insertions are directly targeted by selection in populations successfully coping with environmental stress owing to the TE activity, other neutral insertions will also be transmitted to subsequent generations resulting in the overall copy number increase.

\subsection{Conclusions}
In this paper, we developed a model of TE proliferation mechanism in diploid sexual organisms that face the environmental changes. The model allows to follow the evolution of TE copies and their mutational influenece on the host population.
Our simulations invetigate the interaction between active and inactive copies suggesting that the presence of inactive copies (in the quasi-steady state) is essential to preserve the transposition-selection equilibrium of active copies. Moreover, we suggest that additional mutational power related to the TE activity becomes crucial when the population needs to keep up in the fast or rapidly changing environment.  

When it comes to the further research our main goal is to derive an analitical description of our model and then verify if theoretical calculations are consistent with the computational predictions. However, this goal seems to be highly nontrivial, since we introduce a natural selection and reproduction mechanisms based on the global fitness of the population. Thus, popular apporaches such as evolution of Markov Chains are discarded. 
Additionaly, we would like to explore the influence of other types of environmental change, e.g. fluctuating changes or abrupt changes of large amplitude.
\section{Acknowledgments}
\label{ch:acknowledgments}
This work was partially supported by Polish National Science Center grant 2012/06/M/ST6/00438 and grant POLONIUM: \textit{Matematyczne i obliczeniowe modelowanie ewolucji ruchomych element\'{o}w genetycznych}.
\bibliographystyle{apalike}
\bibliography{paper}
\end{document}